\documentclass[conference]{IEEEtran}

\usepackage{ucs}
\usepackage[utf8x]{inputenc}
\usepackage[T2A,T1]{autofe}
\usepackage{makeidx}
\usepackage{graphicx}
\usepackage{stringenc}
\usepackage[filenameencoding=utf8x]{grffile}
\usepackage{tabularx}
\usepackage{color}
\usepackage{flushend}
\usepackage{listings}
\usepackage{multicol}

\title{Operating the Cloud from Inside Out}
\author{\IEEEauthorblockN{Josef Spillner\IEEEauthorrefmark{1,2}, Andrii Chaichenko\IEEEauthorrefmark{1}, Andrey Brito\IEEEauthorrefmark{2}, Francisco Brasileiro\IEEEauthorrefmark{2}, Alexander Schill\IEEEauthorrefmark{1}}
\IEEEauthorblockA{\IEEEauthorrefmark{1}
Faculty of Computer Science\\
Technische Universität Dresden\\
01062 Dresden, Germany\\
Email: \{josef.spillner,alexander.schill\}@tu-dresden.de, andrii.chaichenko@mailbox.tu-dresden.de}
\IEEEauthorblockA{\IEEEauthorrefmark{2}
Laboratório de Sistemas Distribuídos\\
Universidade Federal de Campina Grande\\
CEP 58429-900, Campina Grande - PB, Brazil\\
Email: \{andrey,fubica\}@dsc.ufcg.edu.br}}

\begin{document}

\maketitle

\begin{abstract}
Virtual machine images and instances (VMs) in cloud computing centres are typically designed as isolation containers for applications,
databases and networking functions. In order to build complex distributed applications, multiple virtual machines
must be connected, orchestrated and combined with platform and infrastructure services from the hosting environment.
There are several reasons why sometimes it is beneficial to introduce a new layer, Cloud-in-a-VM, which acts
as a portable management interface to a cluster of VMs. We reason about the benefits and present our Cloud-in-a-VM
implementation called Nested Cloud which allows consumers to become light-weight cloud operators on demand and
reap multiple advantages, including fully utilised resource allocations.
The practical usefulness and the performance of the intermediate cloud stack VM
are evaluated in a marketplace scenario.
\end{abstract}


\section{Limitations of Public Clouds}

Cloud computing is, for a reason, also called utility computing: all infrastructure, software and data should be
available on demand, at the right scale, and for a price which precisely corresponds to the usage pattern \cite{ibmutility}.
In practice, public cloud providers only offer utility computing at a first approximation.
The on-demand or ad-hoc acquisition is limited by manual sign-up processes which involve credit cards and
personal data; the vertical scalability is coarse-grained and the horizontal scalability subject to customised
applications; and the billing is performed in needlessly large time slots \cite{nestedcloudanalysis}.
Furthermore, the management APIs and graphical user interfaces cannot fulfil all potential requirements on
a scale between full automation and high user-friendliness.

The intertwined nature of cloud services and corresponding management services contributes heavily to this
issue. For multiple reasons, to be detailed later, it would be desirable to introduce an intermediate layer
under the control of the user. As opposed to existing cross-cloud management tools and portals \cite{cloudmanagement},
it would operate directly within the user's allocation.
We envision a Cloud-in-a-VM scenario which, when deployed in a compute cloud, enables the infrastructure service
consumer to become a light-weight provider and overcome the mentioned operational limitations through proper provisioning
and management tools. The VM which contains the intermediate cloud and which operates subordinate application VMs
would be externally indistinguishable from traditional application VMs.

One evident practical requirement for the envisioned Cloud-in-a-VM approach is the pass-through and thus the availability
of hardware virtualisation features on the VM level. The virtualisation extensions of current 64-bit processors
(Intel VT-x/VMX, AMD-V/SVM, ARM AVE) have led to sophisticated nested virtualisation support in several hypervisors
(KVM 1.0+ on Linux 3.2+, VMware vSphere 5.0+ with VHV switch, Xen 4.2+ experimentally).
This means that a hypervisor on the lowest level (L0) can operate a VM (L1) with a hypervisor which can operate
an internal VM (L2) \cite{nestedcloudanalysis}.
Users of cloud computing stacks (Xen Cloud Platform, OpenStack and others) benefit from the pass-through.
The nesting is often limited to specific combinations (e.g. KVM-on-KVM and KVM-on-Xen), but constantly expanding,
and somewhat limited in terms of L2 performance, but constantly improving \cite{intelnestedvirtupdate}.
Other hypervisors (Microsoft
Hyper-V, Oracle VirtualBox) don't support nesting, and hence cloud stacks and infrastructure services on top of them (Microsoft
Azure) do neither. Even when nesting-capable hypervisors are employed by a cloud provider, nesting may be switched off
(e.g. Amazon EC2 on Xen) for technical or business reasons. Our Cloud-in-a-VM approach assumes that nesting is available
in the hypervisor which runs the VM instances in a public cloud. In other words, once the pass-through limitation
is overcome, all other mentioned limitations can be overcome as well.



\section{Benefits of Cloud-in-a-VM}

The recent terms {\it Meta Cloud}, {\it Nested Cloud}, {\it Cloud of Clouds} and {\it Intercloud} all describe sharpening trends
towards higher operational control across cloud providers' technical and business boundaries \cite{jitfederation,commcloud}.
In particular, nested clouds are known to offer benefits over a flat VM-to-provider deployment model \cite{hvcrb}.
These benefits encompass a higher degree of utilisation, more convenient deployment of a large number of already
interconnected VMs, economic gains through re-purposing otherwise idle and already paid-for resources, and
unified interaction with consistent provider-independent user interfaces.

\begin{itemize}
\item Higher degree of utilisation. Often, VM allocations in the public cloud are subject to coarse-grained
scaling and duration. With a Cloud-in-a-VM, these limits can not be overcome, but the allocation can be more efficiently
utilised by re-purposing the unused slices.
\item Economic gains. In combination with the higher degree of utilisation, re-selling otherwise unused slices
grants an income which in some cases may even cumulatively offset the initial purchasing cost. For instance,
a large VM instance can be reserved early on for a good price and later in parts be re-sold as subordinate spot instances.
\item More convenient deployment. When large numbers of VMs need to be deployed from existing instances,
it becomes easier to just migrate (virtualise) their host machines instead of migrating each VM on its own.
\item Consistent user interfaces. Instead of having to learn and use a different user interface for each cloud
provider, a distributed Cloud-in-a-VM offers a unified management interface with at least a common subset of
operational controls such as restart, rescale and snapshot creation.
\end{itemize}

\begin{figure}[h]
\center
\includegraphics[width=\columnwidth]{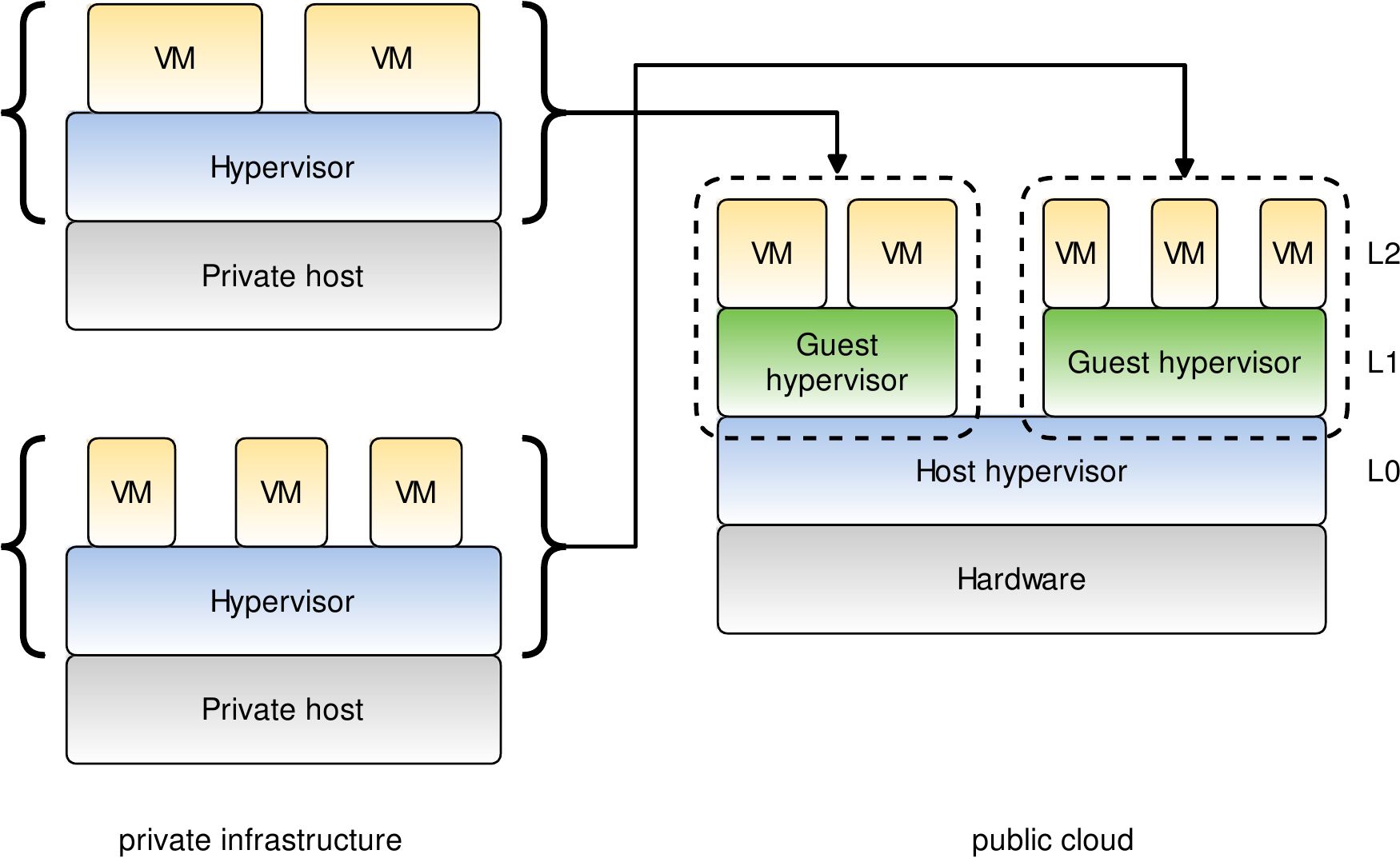}
\caption{Principal system topology involving nested virtualisation}
\label{fig:cloudstacks}
\end{figure}

Nested clouds are typically designed using nested virtualisation to avail oneself of hardware acceleration.
This concept is outlined in Fig. \ref{fig:cloudstacks} using the following example: Two physical hosts which
run two and three VMs, respectively, are subject to being virtualised. Instead of allocating five new VMs
and their dependencies, the virtualisation is performed at the host level, thus moving the operating system and hypervisor
to an application level on the target host which already runs a hypervisor.

However, so far, nested clouds have only been superficially analysed, primarily on the underlying virtualisation
level \cite{nestedcloudanalysis,cloudvisor}.
No implementation of a usable nested cloud is known to the authors, and no insight-delivering measurement results are available.
Therefore, it is our aim to introduce {\it Nested Cloud} as a prototypical representative of a nested Cloud-in-a-VM
and evaluate it in the context of a market-driven cloud resource service economy.

\section{Nested Cloud, a High-Utility Cloud-in-a-VM}

Nested Cloud is a VM which contains a complete dynamic cloud computing management environment. Its functions include
the start and stop of service instances which run as sub VMs as well as their (re-)configuration in terms of variable
numbers of CPU cores, CPU priorities, amount of RAM, available disk space and network adapters.
The Nested Cloud VM is to be deployed into a cloud provider's IaaS service following the same procedure as for application VMs.

\begin{figure}[h]
\center
\includegraphics[width=\columnwidth]{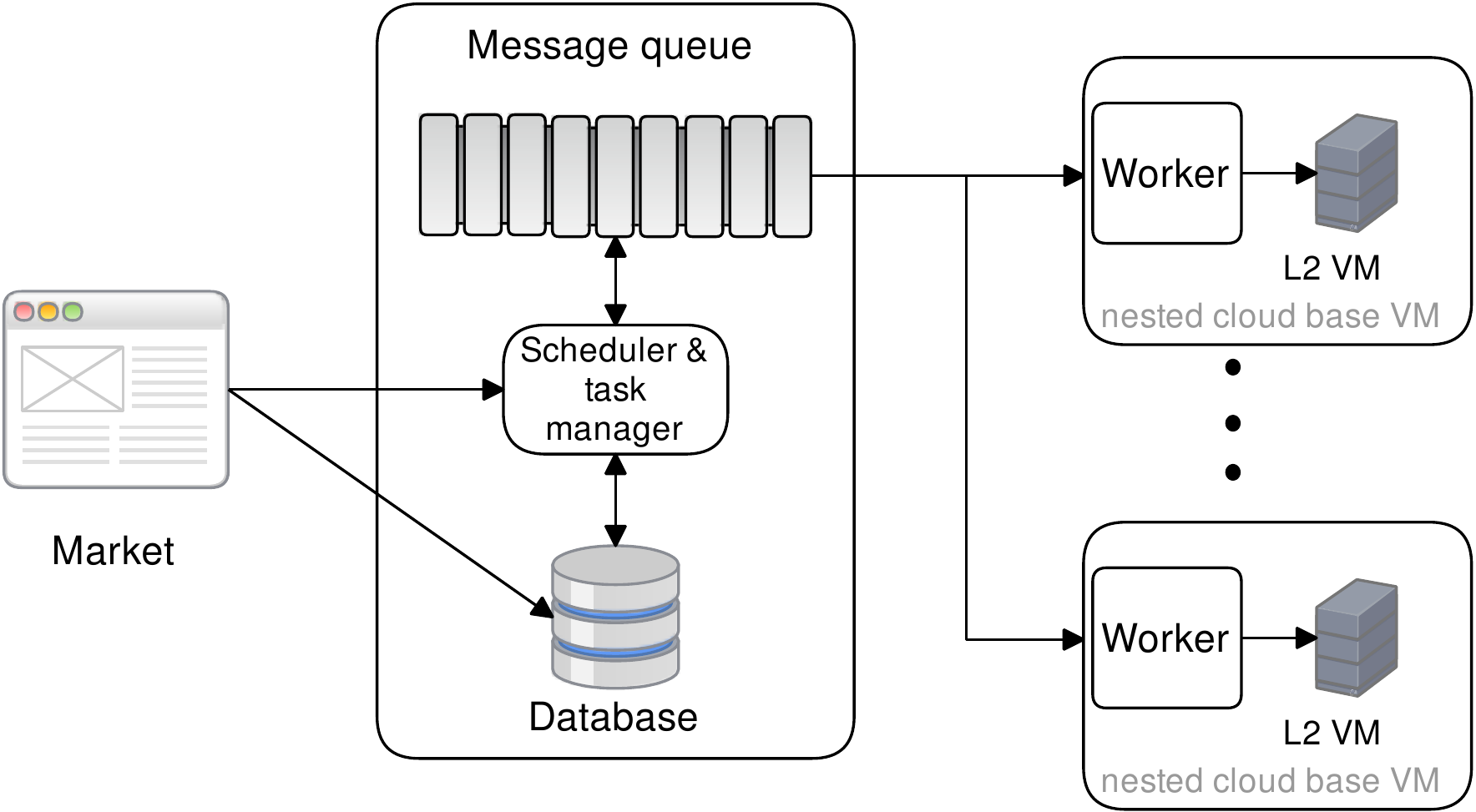}
\caption{The satellite worker model of Nested Cloud}
\label{fig:backend}
\end{figure}

Fig. \ref{fig:backend} conveys the general structure of Nested Cloud. It contains a central node with a database,
a fault-tolerant message queue and a scheduler, surrounded by satellite workers which act on a higher level of virtualisation.
The workers attach to the message queue and receive commands in an event-based programming style.
Launching a VM shall serve as example command. On arrival of the command signal, an XML definition containing the VM
configuration is being created. It contains the unique VM UUID, the human-readable name, the resource definitions as mentioned
in the previous paragraph and a reference to the bootable image file.
The set of commands also contains the allocation of resources scheduled at a future time, the rescaling of resources
and the retrieval of system status information. In addition, block storage management commands are available. Their set contains
the allocation, resizing and removal of storage areas. The areas are formatted with a filesystem and can be used to store
instance snapshots or auxiliary data from the applications themselves.

The implementation of Nested Cloud makes use of the recursively nested virtualisation (RecVirt)
project\footnote{RecVirt: http://gitorious.org/recvirt}.
RecVirt generates custom Linux kernels and Kernel Virtual Machine (KVM) setups and wraps them using
Debootstrap\footnote{Debootstrap: http://wiki.debian.org/Debootstrap}
and additional Debian package lists into a minimised Linux system structure to yield a bootable VM disk image.
As extension to RecVirt, it sets up a minimal cloud computing environment: An NFS mountpoint to access application disk
images and another one to store block images which serve as complementary data disks.

The workers are implemented in Python with libvirt to control virtual machines and the Celery task/message queue system.
All scripts to produce Nested Cloud are provided publicly as part of the SPACE-Cloud
repository\footnote{SPACE-Cloud Repository: http://serviceplatform.org/wiki/Cloud}.

By design, the system does not include any off-the-shelf cloud computing
stack to keep it lean and manageable for research.
These stacks not only offer compute and block storage services, but also a vast amount of additional
functionality starting from durable object storage, databases, event processing, load balancing and automatic
scaling \cite{cloudtaxonomy}. Suitable open source stacks are OpenNebula, OpenStack, Eucalyptus, Nimbus, Cloud Foundry
and the Xen Cloud Platform \cite{cloudstacksurvey,cloudstackcritique}.
We acknowledge the need to integrate the
backend execution into such stacks in the future to attract practicioners and commercial deployments.
From a technical point of view, most of them can be configured to run on a hypervisor like KVM so the general
observations about nested clouds remain valid.


\section{Nested Cloud in a Cloud Resource Economy}

\begin{figure}[h]
\center
\includegraphics[width=\columnwidth]{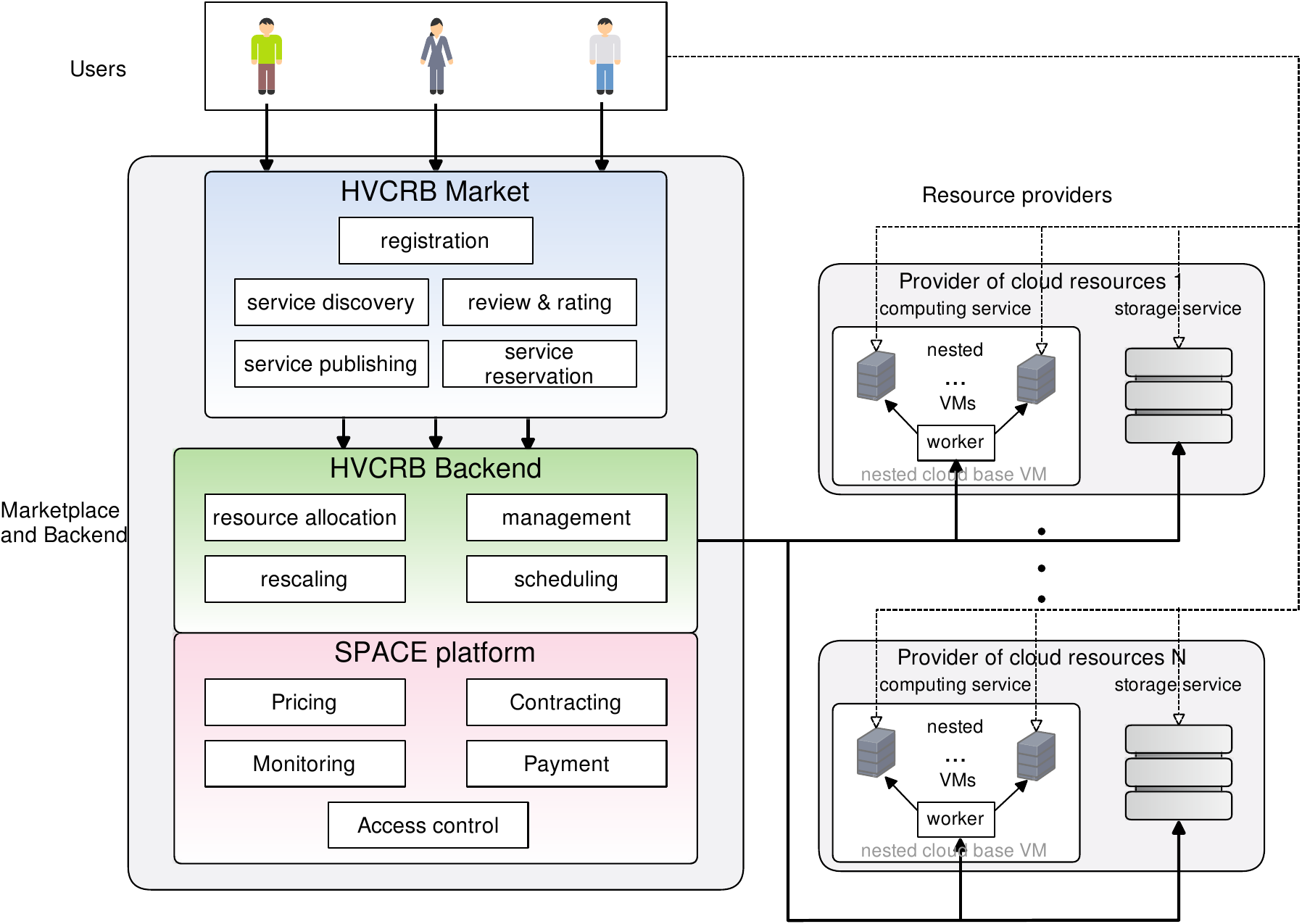}
\caption{Overall intended architecture in which Nested Clouds and application VMs are managed through a resource market}
\label{fig:hvcrbsystem}
\end{figure}

Service economies are spaces in which supply and demand of services are regulated through open marketplaces.
Especially for cloud resource services, markets need to compete with flexible options to reach a critical
mass of users. A market which offers the technical flexibility and economic advantages of Nested Cloud along with other expected features
(real-time updates, high scalability, user friendliness) is going to raise the bar for future service market
development. Hence, we position HVCRB, the Highly-Virtualising Cloud Resource Broker \cite{hvcrb}, as prototype for such
a market. It consists of a marketplace frontend and one or more Nested Cloud backends located at various distributed infrastructure
operators. Along with the market, the SPACE service platform\footnote{SPACE Service Platform: http://serviceplatform.org/projects/space}
runs as technological foundation for all service management and execution aspects. Fig. \ref{fig:hvcrbsystem} gives
an overview about the architecture.

\begin{figure}[h]
\center
\includegraphics[width=\columnwidth]{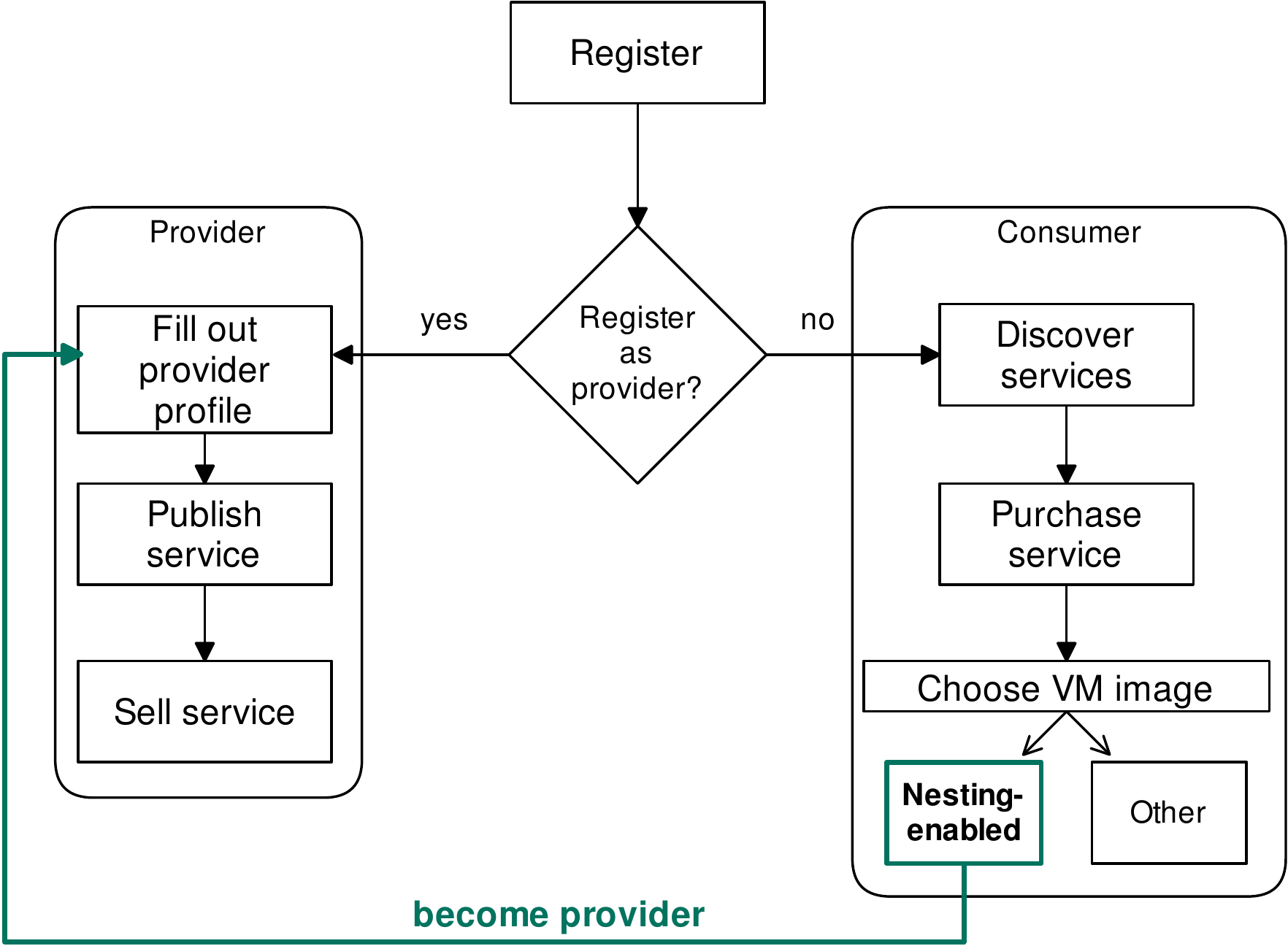}
\caption{Workflow for market users who assume a provider and/or consumer role}
\label{fig:workflow}
\end{figure}

Users access the marketplace to offer or consume services. The workflow is depicted in Fig. \ref{fig:workflow}.
They can assume both roles, consumer and/or provider, on demand.
The distinguishing feature here is the connection from consumer to (casual) provider through the use
of the Nested Cloud VM. In order to perform the transition, the consumer only needs to specify additional
information which for legal or business reason need to be known for all providers. This might be the company
name, the tax number, bank account information or the postal mail address.

\begin{figure}[h]
\center
\includegraphics[width=\columnwidth]{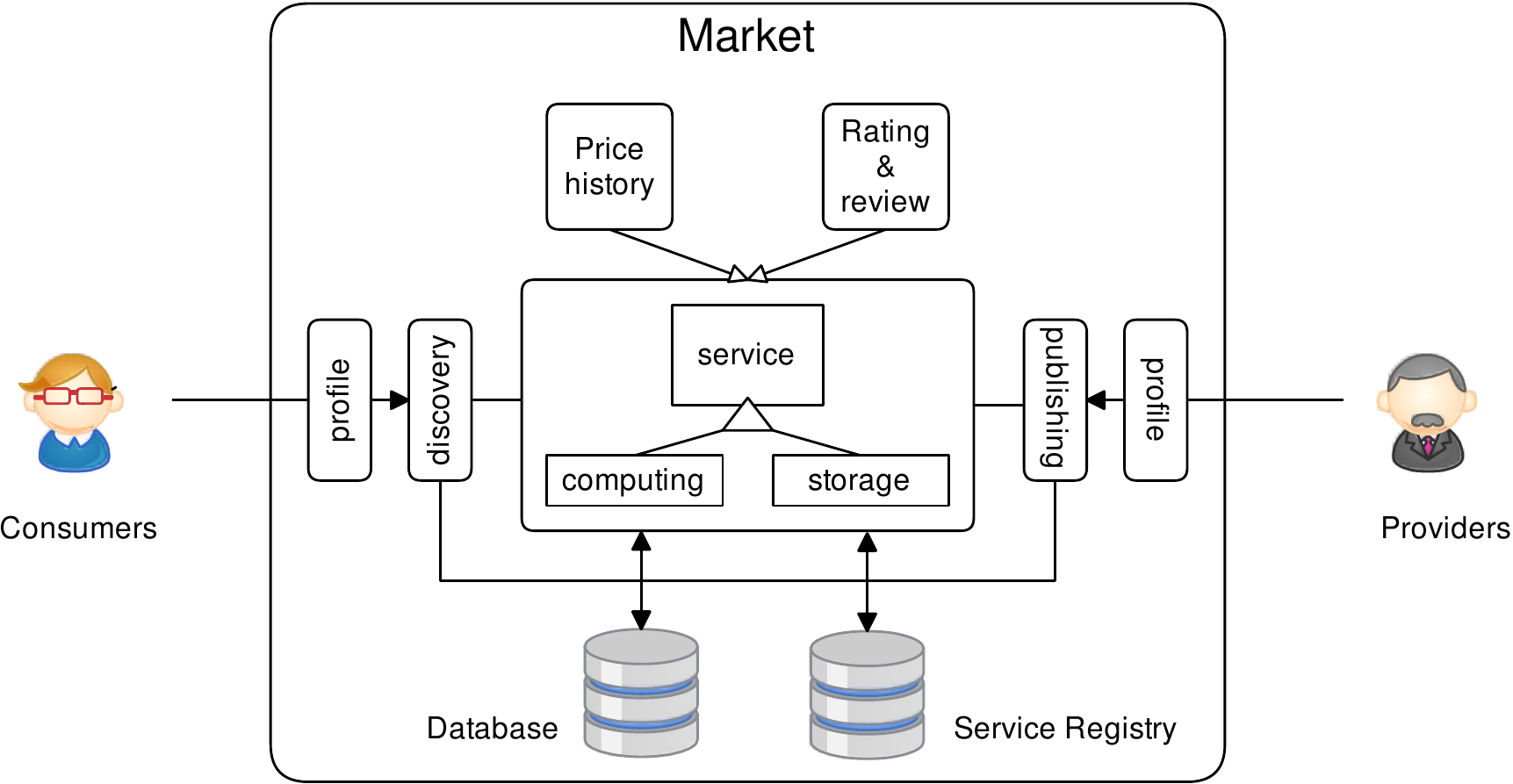}
\caption{Characteristics of a cloud resource market for compute and storage services}
\label{fig:frontend}
\end{figure}

Fig. \ref{fig:frontend} complements the workflow by giving an overview about the features and characteristics of the marketplace.
One can relate its two content types, compute and storage services, and the two roles, cloud resource service provider and consumer.
Other cloud services are not considered here, although can be trivially added through a registration of the corresponding domain
ontologies into the service registry.

\begin{figure}[h]
\center
\includegraphics[width=\columnwidth]{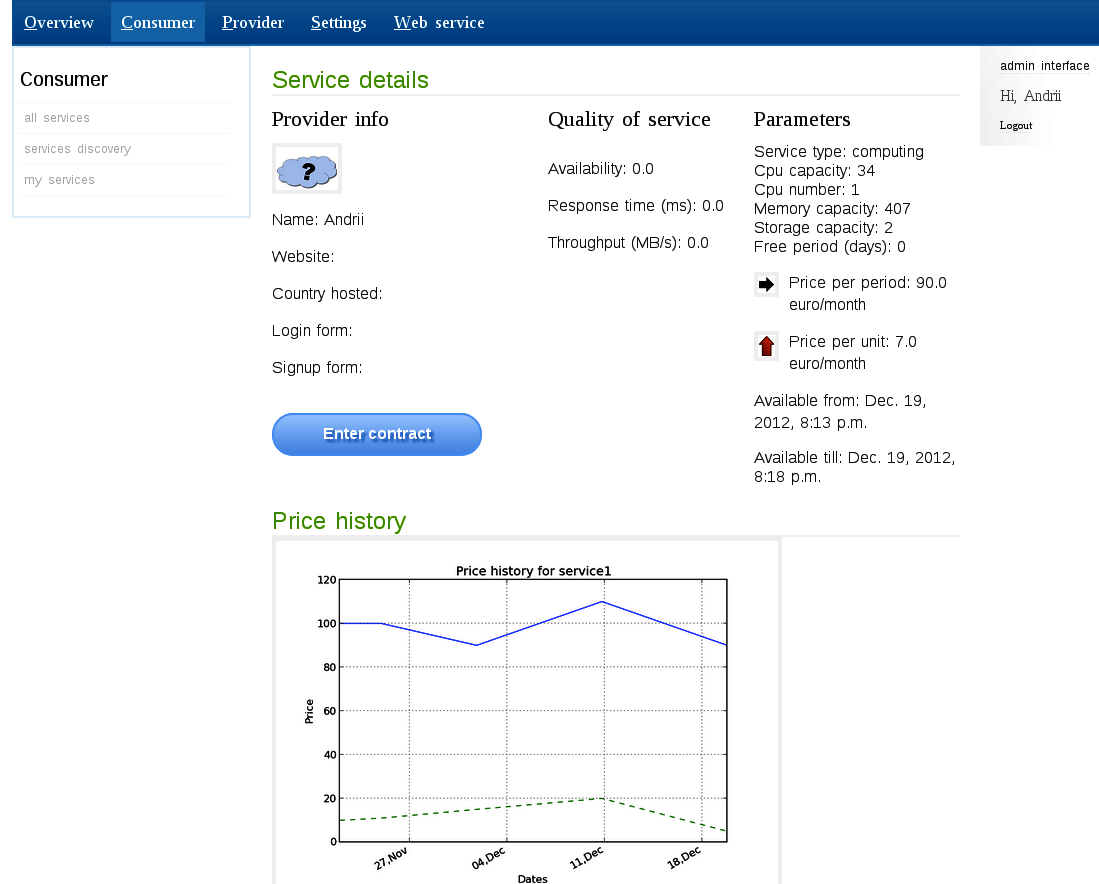}
\caption{Screenshot of HVCRB, a spot market for trading and controlling cloud resources}
\label{fig:marketshot}
\end{figure}

The marketplace is realised as a Django web application which connects to a relational database
(SQlite or MySQL) and to the service registry ConQo \cite{xaasregistry}.
A screenshot of a compute service information page from the marketplace is shown in Fig. \ref{fig:marketshot}.
The service, to which a contractual does not exist yet, has presumably been selected beforehand.
Along with a price history information widget, the page lists general about the service and its provider
as well as service quality information. Once a contract is negotiated and established, the allocation
can be configured, started, stopped and otherwise be interacted with.

In essence, HVCRB demonstrates the feasibility to build marketplaces and fuel service economies with
distinguished features such as the ability for consumers to choose the Nested Cloud VM and micro-manage
all further allocations through it.


\section{Nested Cloud Experimental Results}

\begin{figure}[h]
\center
\includegraphics[width=\columnwidth]{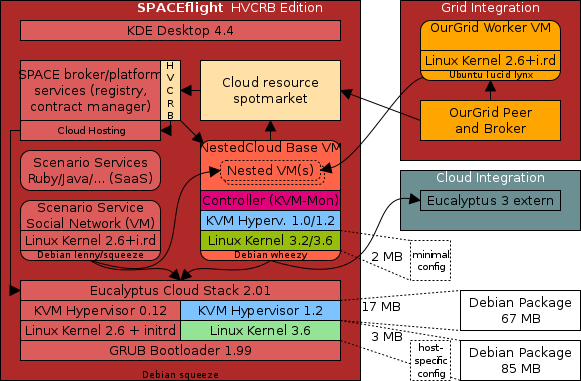}
\caption{Integration of Nested Cloud into SPACEflight by way of introducing a dual stack}
\label{fig:integrateddemo}
\end{figure}

Contrary to the previous nested virtualisation overhead analysis \cite{nestedcloudanalysis},
we have evaluated Nested Cloud in a real application scenario with a web application running in an
L2 VM inside a cloud realised by a lower-level L1 Nested Cloud VM.
For reasons of preservation and reproducibility of the experiments, both VM levels and the HVCRB spot market are integrated into SPACEflight,
an open source mini-ecosystem for service and cloud technologies\footnote{SPACEflight: http://serviceplatform.org/wiki/SPACEflight}.
Fig. \ref{fig:integrateddemo} explains the dual-stack setup with the default software components on the left
side and custom latest kernel and KVM configurations beneath Nested Cloud on the right side.

\begin{figure}[h]
\center
\includegraphics[width=\columnwidth]{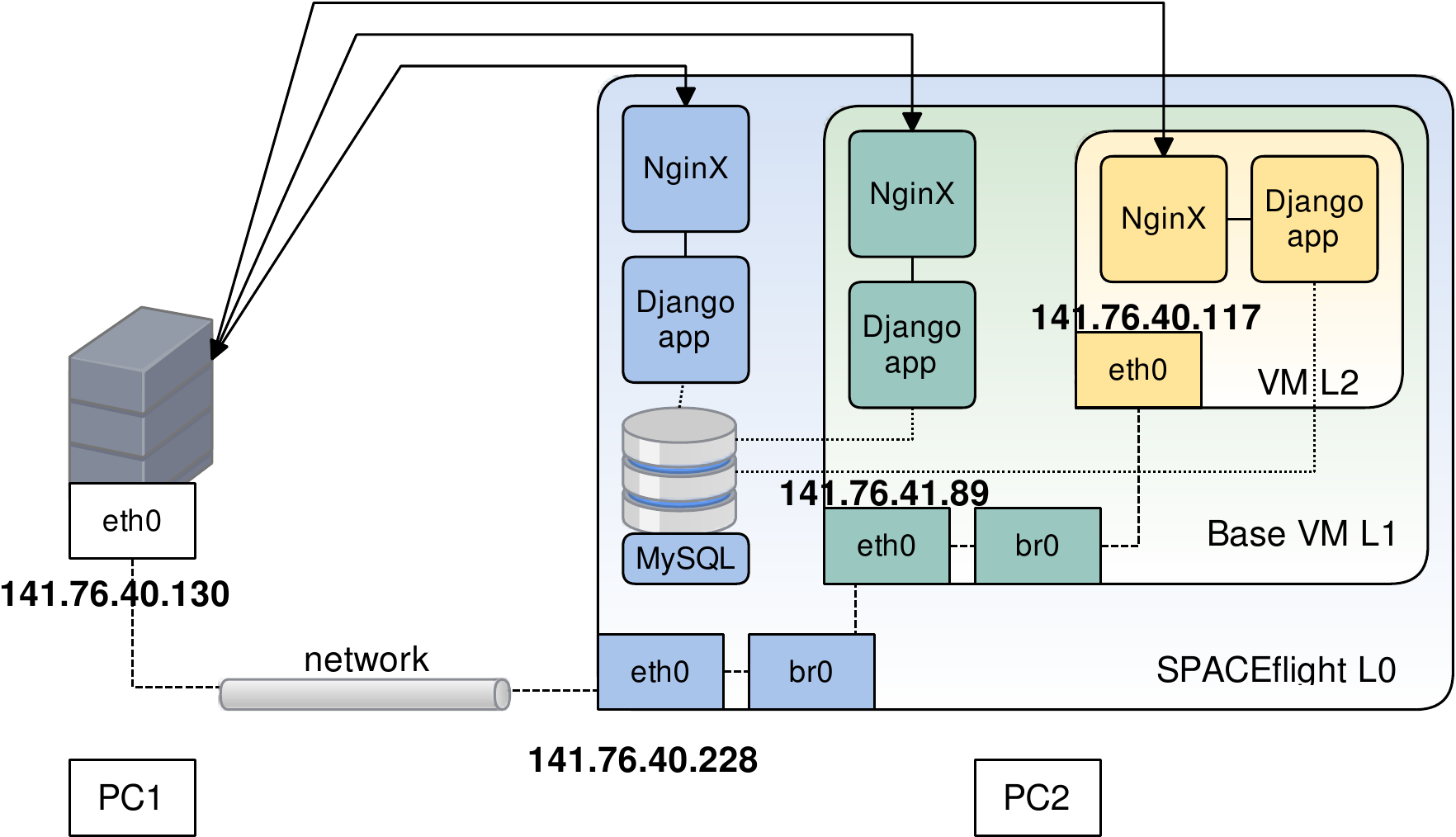}
\caption{Testbed configuration with two nodes}
\label{fig:testbed}
\end{figure}

The experiment testbed consists of two computers as shown in Fig. \ref{fig:testbed}. The
client to the left sends web application requests and measures the response times, whereas
the one to the right runs the nested cloud. On each virtualisation level, there is an instance of the
NginX web server as reverse proxy to
a Django web application which is attached to a MySQL database through an relational-object
mapping. The test request selects 10 rows, converts them to objects and fills an HTML page template
with them. In periods of 3 minutes each, 64 concurrent users are simulated with random
request intervals. The response times are measured and the overhead is calculated based on
these probes.

The hardware and software configuration of the two computers has been specified as follows.

\begin{itemize}
\item PC 1: CPU Intel Core2 Duo P8600 (2.4 GHz, two cores w/ 3 MB L3 cache), RAM 4 GB,
Linux kernel 3.6.3, ConQo service registry, Apache2 web server with mod\_wsgi.
\item PC 2: CPU Intel Core i5 M520 (2.4 GHz, two cores (four virtual) w/ 3 MB L3 cache), RAM 4 GB,
Linux kernel 3.5, NginX HTTP proxy.
\item Both PCs: KVM 1.2, libvirt, MySQL 5.1.49.
\end{itemize}

\begin{figure}[h]
\center
\includegraphics[width=\columnwidth]{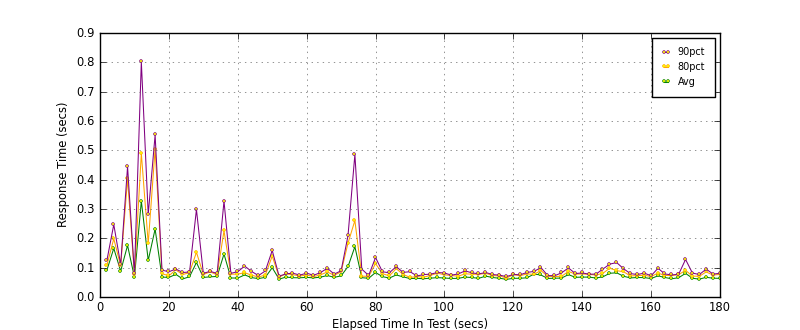}
\includegraphics[width=\columnwidth]{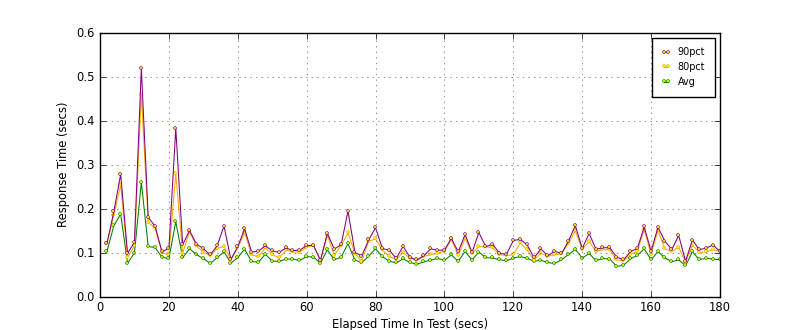}
\includegraphics[width=\columnwidth]{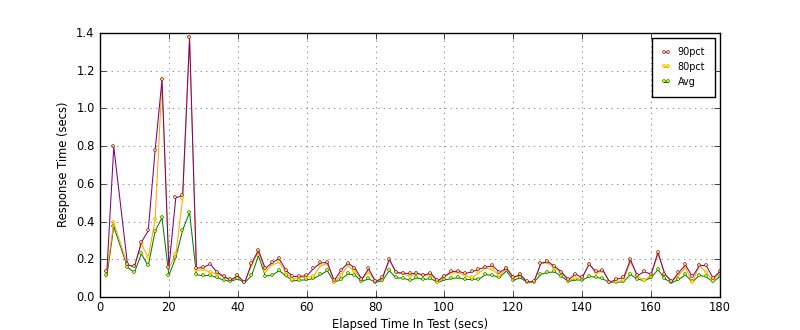}
\caption{Measured response times across all virtualisation levels: L0 (host) in the first,
L1 (Nested Cloud) in the second and L2 (application VM) in the third row}
\label{fig:levelresponsetimes}
\end{figure}

The results in Fig. \ref{fig:levelresponsetimes} clearly show calibration
spikes in the first 30-40 seconds which then flatten due to operating system, web server and database caching effects.
They also show the influence of the virtualisation and (albeit nigligible) nested cloud
management overhead such as rescaling commands which work in real-time.

In Tab. \ref{tab:statistics}, the key metrics are compared. It becomes apparent
that the L1 overhead over L0 is 17.07\% and L2 over L1 is 52.44\%, which is very different
from the values presented in \cite{nestedcloudanalysis} and related nested virtualisation
literature. This discrepancy can be explained by the real-world nature of the application
which utilises all resources (processor, memory and I/O) at once. Furthermore, non-optimised
and CPU-intensive VirtIO drivers contribute to the results \cite{intelnestedvirtupdate,virtioperformance}.

\begin{table}[h]
\center
\caption{HTTP request duration, in seconds}
\label{tab:statistics}
\begin{tabular}{|l|lll|}
\hline
Nesting level & Avg ø & 80\%-ile & 90\%-ile \\
\hline
L0            & 0.082 & 0.081 & 0.098 \\
L1            & 0.096 & 0.109 & 0.128 \\
L2            & 0.125 & 0.144 & 0.181 \\
\hline
\end{tabular}
\end{table}


\section{Conclusion}

Nested Cloud, a realisation of the Cloud-in-a-VM idea, can be deployed as instance into public or private IaaS
providers to achieve higher control about the allocation and scheduling of application VMs. Currently, the two
main obstacles for practical operation are the relatively high run-time overhead for practical use cases and
the simplified assumptions about network resource sharing, e.g. through elastic IP addresses.
We encourage further work on these two aspects to increase the chances for portable provider-independent
resource markets and Cloud-in-a-VM management solutions.


\section*{Acknowledgements}

This work has received funding under project number 080949277 by means of the European Regional Development
Fund (ERDF), the European Social Fund (ESF) and the German Free State of Saxony.
It has also been funded by a scholarship from the Brazilian National Council for Scientific and Technological Development - CNPq (PDJ 159716/2011-0).

\bibliographystyle{IEEEtran}
\bibliography{operating-the-cloud}

\end{document}